# Show by a theoretical and experimental argument that potassium atoms possess a permanent electric dipole moment based on the orbital angular momentum

Pei-Lin You

Institute of Quantum Electronics, Guangdong Ocean University Zhanjiang 524025 China.

The permanent electric dipole moment (EDM) of the ground state of potassium has been found by measuring the electric susceptibility. We find $d_K=[1.58\pm0.19(stat)\pm0.13(syst)]\times10^{-8}$e.cm and the induced EDM, $d_{ind} < 1.5\times10^{-16}$e.cm, can be neglected. The experimental K material with purity 99.95 % is supplied by Strem Chemicals Co. USA. This paper shows that the permanent EDM is based on the orbital angular momentum (where the spin is irrelevant), and neither space inversion nor time reversal is a symmetry operation for an alkali atom. In fact, the famous Runge-Lenz vector **M** is simply the permanent EDM vector of an alkali atom. As early as 1926, W. Pauli calculated the energy levels of the hydrogen using the vector **M**. Therefore, our work may fall into the category of the most exciting discoveries during the past few decades.

**1. Introduction**     The existence of a finite permanent electric dipole moment (EDM) of an atom or particle would violate time reversal symmetry (T), and would also imply violation of the combined charge conjugation and parity symmetry (CP) through the CPT theorem [1-4]. The currently accepted Standard Model of Particle Physics predicts that the dipole moments of an atom are unobservable; therefore, EDM experiments are an ideal probe for new physics beyond the Standard Model. Experiments to search for an EDM of an atom began many decades ago. As early as 1968, M. C. Weisskopf, the Director of CERN, reported an upper limit to the EDM of the cesium atom ($-d_{Cs}< 3.7\times 10^{-22}$e.cm) [4]. "The importance of these experiments lies in the fact the observation of an EDM in an atomic system of well-defined angular momentum would be direct evidence for a violation of both parity and time-reversal invariance." From then on, many brilliant physicists have pursued the subject. Experimental searches for EDMs can be divided into three categories: search for the neutron EDM (the new result is $d_n< 2.9\times10^{-26}$e.cm)[2], search for the electron EDM utilizing paramagnetic atoms, the most sensitive of which is done with Tl atoms (the result is $d_e=[1.8\pm1.2 (stat)\pm1.0 (syst)]\times10^{-27}$e.cm) [3], and search for the EDM of diamagnetic atoms, the most sensitive of which is done with $^{199}$Hg (the new result is $d(Hg)=[0.49\pm1.29(stat)\pm0.76 (syst)] \times 10^{-29}$e.cm) [1]. In all experiments, the microcosmic Larmor precession frequency of individual particles were measured based on nuclear spin or electron spin. The search for an EDM involves measuring the precession frequency of the particle in parallel electric and magnetic fields and looking for a change of this frequency when the direction of **E** is reversed relative to that of **B**. Despite the relentless search for a non-zero EDM, lasting more than 50 years, no conclusive results have been obtained [1-4].

Up to now, physicists commonly believed that if a particle has spin zero (or the spin is irrelevant), the permanent EDM of the particle must be zero [1-4,13]. **In fact, the idea is an untested hypothesis that deserves exhaustive examination.** In 2002 we reported a lower limit to the permanent EDM of an Rb atom by measuring the electric susceptibility, $d_{Rb}\geq 8.6\times10^{-9}$e.cm[5]. If nearly all the dipoles in a gas turn toward the direction of the external field, this effect is called the saturation polarisation. In 2004 we reported that the saturation polarisation of Rb vapour was observed [15]. After nine years of intense research, this paper will report that the saturation polarisation of K vapour was observed and the permanent EDM of a K atom has been measured accurately. The main experimental apparatus, two cylindrical capacitors containing K vapour, were made in the Department of Physics of Peking University, and above results repeated in their laboratory.

**2. Prove the ground state alkali atom may have a permanent EDM based on the orbital angular momentum.**

It is well known that the ground state of hydrogen is non-polar atom (d=0). The linear Stark effect shows that a hydrogen atom of the first excited state has large permanent EDM, $d_H=3e\,a_0=1.59\times10^{-8}$e.cm ($a_0$ is the Bohr radius) [6]. L.D. Landay once stated that "The presence of the linear effect means that, in the unperturbed state, the hydrogen atom has a dipole moment"[7]. Therefore, the excited state of hydrogen is polar atom. We notice that the permanent EDM of the Hg atom based on nuclear or electron spin is only $10^{-29}$e.cm. It is apparent that such a large permanent hydrogen EDM cannot be based on nuclear spin or electron spin! Its radius, $r_H = 4\, a_0 = 2.12\times 10^{-8}$ cm, is almost the same as $^{199}$Hg ($r_{Hg} =1.51\times 10^{-8}$ cm) [8], but their EDMs differ by twenty-one orders of magnitude! In modern quantum theory, the electrons are described as charge clouds rather than as orbiting particles. How do we explain this inconceivable discrepancy? The existing theory cannot answer the problem! However, a hydrogen or hydrogen-like atom may have a large permanent EDM according to the Sommerfeld theory (1916). In this theory, the electron moves along a quantization elliptic orbit. We can draw a straight line perpendicular to the major axis of the elliptic orbit through the nucleus in the orbital plane. The line divides the elliptic orbit into two parts that differ in size. According Kepler's law of equal areas, the average distance between the moving electron and the nucleus is larger and the electron remains in the larger area longer than in the smaller area. As a result, the time-averaged value of the electric dipole moment over a period is nonzero for the atom.

We investigate the classical Kepler problem. The Hamiltonian in relative coordinates is[6,9]

$$H=\mathbf{p}^2/2\mu - \kappa/r \qquad (1)$$

where μ is the reduced mass and $\kappa=Ze^2$ for the hydrogen-like atom. The bound solution of the classical orbit problem is an ellipse with semi-major axis *a* that is equal to half the distance from perihelion *P* to aphelion *A*, and with eccentricity ε that is equal to $(a^2 - b^2)^{1/2}/a$, where *b* is the semi-minor axis. The total energy E and the orbital angular momentum L= $\mathbf{r} \times \mathbf{p}$ are two constants of the motion. The rotational symmetry of H is not enough to require the orbit to be closed. This fact suggests that there is some quantity, other than H and **L**, that is a constant of the motion and that can be used to characterize the orientation of the major axis in the orbital plane. Therefore, we look for a constant vector **M** which we expect to lie along the major axis, pointing from the focus *O* of the ellipse to *P*. Such a vector has been known for a long time and is called the Runge-Lenz vector **M** or eccentricity vector **Γ** [6,10].

$$\mathbf{M} = \mathbf{p}\times\mathbf{L}/\mu -\kappa\, \mathbf{r}/r \quad \text{or} \quad \mathbf{\Gamma}= \mathbf{M}/\kappa=\mathbf{p}\times\mathbf{L}/\kappa\mu -\mathbf{r}/r \qquad (2)$$

It is easily seen to be a constant of the motion, to have magnitude κε (or ε) directed from *O* to perihelion *P* [9,10].

$$\mathbf{M} = \kappa\varepsilon\, \mathbf{r}_p/r_p \quad \text{or} \quad \mathbf{\Gamma}= \varepsilon\, \mathbf{r}_p/r_p \qquad (3)$$

where $\mathbf{r}_p$ is the vector directed from *O* to perihelion *P*. To treat the hydrogen atom quantum mechanically, we must to replace the classical functions by operators, which can be done easily for **r**, **p**, and **L**. Because $\mathbf{p}\times\mathbf{L} \neq -\mathbf{L}\times\mathbf{p}$, Eq. (2) does not define a Hermitian quantity. We therefore redefine **M** as a symmetrized expression [6]:

$$\mathbf{M} = (\mathbf{p}\times\mathbf{L} - \mathbf{L}\times\mathbf{p})/2\mu -\kappa\, \mathbf{r}/r \qquad (4)$$

By considering the commutation relations for **r** and **p** we can show that

$$[\mathbf{M}, H]=0, \qquad \mathbf{L}\cdot\mathbf{M}=\mathbf{M}\cdot\mathbf{L}=0, \qquad \mathbf{M}^2= 2H(\mathbf{L}^2+\hbar^2)/\mu + \kappa^2 \qquad (5)$$

One of the world's greatest theoretical physicists, Wolfgang Pauli calculated the energy levels of the hydrogen atom in 1926 using Eq. (4) and (5): $E= -\mu\kappa^2/2n^2\hbar^2$. Pauli's approach is equivalent to regarding the three components of **M** as generators of some infinitesimal transformations [6].

On the other hand, J.J. Mestayer et al demonstrated a protocol to create localised wave packets of potassium atoms in 2008. "This figure shows a typical classical trajectory of an electron in a highly elliptical orbit oriented along the *x* axis when potassium atoms are suddenly exposed to the transverse field." [11]

In quantum mechanics, **alkali atoms with only one valence electron in the outermost shell can be described as hydrogen-like atoms**.[12] The quantum numbers of the ground state of alkali atoms are n≥2 rather than n=1(this is 2 for Li, 3 for Na, 4 for K, etc.), as the excited state in hydrogen.

L.I. Schiff wrote that "A charged particle with spin operator **S** will possess an electric dipole moment operator **d**=μ**S**, where μ is a numerical constant" [6]. In a similar way, a ground state alkali atom with the operator **M** will possess a permanent EDM operator:

$$\mathbf{d} = \gamma\, \mathbf{M}= e\, \mathbf{r} \quad \text{or} \quad \mathbf{d} = \gamma\, \mathbf{\Gamma}/\kappa= e\, \mathbf{r} \qquad (6)$$

where γ =r/Zeε is a constant, and **r** is a constant vector pointing from the moving centre of the valence electron to the nucleus and in the same direction as $\mathbf{r}_p$. Obviously, if a particle has no spin (or the spin is irrelevant), the particle may have a permanent EDM based on the orbital angular momentum **L**. The Hamiltonian for this particle in any electric field **E** contains the interaction term - **dE**.

The Runge-Lenz vector **M** or eccentricity vector **Γ**—is essentially, the permanent EDM vector of an alkali atom. This statement may fall into the category of the most exciting discoveries during the past few decades.

**3. Show that neither space inversion nor time reversal is a symmetry operation for an alkali atom**

The reversal in time of a state represented by the wave function $\Psi_\alpha$ changes it into $\Psi_{\alpha'}$, which develops in accordance with the opposite sense of progression of time. In this new state, the signs of momenta and angular momenta are reversed, but other quantities are unchanged. Time reversal is described by a time-independent operator O(T), which is defined by the equation

$$O(T)\Psi_\alpha(\mathbf{r},t)= \Psi_{\alpha'}(\mathbf{r},t) \qquad (7)$$

It is plausible to expect that the norms of states and the absolute value of the inner products of two states are unchanged by time reversal. We therefore assume that O(T) is antiunitary and write it in the form[6,9,13].

$$O(T)=UK \qquad (8)$$

where the complex conjugation operator K is by definition: $K\Psi=\Psi^*$, U is unitary and $O(T)O(T)^{-1}= O(T)^{-1}O(T) =1$. Note that **J=L+S**, where **L**, **S**, and **J** are the orbital, spin and total angular momentum operator, respectively.

For a zero angular momentum particle:   $U=1$,    $O(T)\Psi(\mathbf{r},t)= \Psi^*(\mathbf{r},-t)$ (9)

For a particle with angular momentum:

$$U=\exp(-i\pi J_y/\hbar), \qquad O(T)\Psi(\mathbf{r},t)= \exp(-i\pi J_y/\hbar)\Psi^*(\mathbf{r},-t)= U_R(\pi)\Psi(\mathbf{r},t) \qquad (10)$$

where $U_R(\pi)$ is the rotation operator and $\exp(-i\pi J_y/\hbar)$ is a unitary transformation that rotates **J** through π radians about the y axis and thus, transforms $J_x$ into $-J_x$ and $J_z$ into $-J_z$. When dealing with **J**, we shall for definiteness always choose a representation in which $\mathbf{J}^2$ and $J_z$ are diagonal. When a particle has no spin (or the spin is irrelevant), **S**=0, and **J=L**. Because **d** = e**r** and the operator **r** is unchanged under time reversal, it follows that **d** satisfies [13]

$$O(T)\mathbf{d}O(T)^{-1}= +\mathbf{d} \qquad (11)$$

Let $\Phi_m$ denote the state of the particle and its projection of the angular momentum on the z-axis equal to *m*. Now consider the expectation value of this equation with respect to the state $O(T)\Phi_m$. Noticing that $(O(T)\chi, O(T)\varphi)_t = (\chi,\varphi)^*_{-t}=(\varphi,\chi)_{-t}$ [13], and the Hermitian of the operator **d**: $(\Phi_m, \mathbf{d}\Phi_m) = (\mathbf{d}\Phi_m, \Phi_m)$, we have

$(O(T)\Phi_m, \mathbf{d}O(T)\Phi_m)=(O(T)\Phi_m,O(T)\mathbf{d}O(T)^{-1}O(T)\Phi_m)=(O(T)\Phi_m,O(T)\mathbf{d}\Phi_m)= (\mathbf{d}\Phi_m, \Phi_m) =(\Phi_m, \mathbf{d}\Phi_m)$    (12)

For a particle with angular momentum, equation (10) is the condition the operator O(T) must satisfy under time reflection, i.e. O(T) acting on a state has the same effect as a rotation by π[13]. Therefore we have

$(O(T)\Phi_m, \mathbf{d}O(T)\Phi_m)=(U_R(\pi)\Phi_m, \mathbf{d}U_R(\pi)\Phi_m) = (\Phi_m, U_R(\pi)^{-1}\mathbf{d}U_R(\pi)\Phi_m)= -(\Phi_m, \mathbf{d}\Phi_m)$,    (13)

because a rotation by π changes the sign of the vector operator **d**[13]. Thus

$$(\Phi_m, \mathbf{d}\Phi_m) = -(\Phi_m, \mathbf{d}\Phi_m) = 0 \qquad (14)$$

Evidently, time-reversal invariant leads directly to the vanishing of the permanent EDM of any atom or particle. Finally, we note that **d** is a polar vector that does change sign on space inversion (parity transformation). The interaction (- **dE**) changes sign when **r** is replaced by (-**r**), and the transformation affected the Hamiltonian of the atomic system. Thus we have clearly proven that neither space inversion nor time reversal are symmetry operations for an alkali atom with the operator **d**, even when no external electric field is present(see Ref.[6] P243,problems19.)

**4. How can we separate the permanent and induced EDM of an atom or molecule?**

We can explain the magnitude of dielectric effects based on simple models of the atomic or molecular dipole moments. The electric susceptibility is defined as $\chi_e =C/C_0–1$, where $C_0$ is the vacuum capacitance and C is the capacitance of the capacitor filled with the material. When atoms are placed in an electric field, they become polarised, acquiring *induced* electric dipole moments in the direction of the field. On the other hand, many molecules do have *permanent* EDM, called polar molecules. Note that the electric susceptibility caused by the orientation of polar molecules is inversely proportional to the absolute temperature, whereas the induced electric susceptibility due to the distortion of electronic motion in atoms or molecules is temperature independent. **This difference in temperature dependence offers a means of separating the two types of EDM experimentally.**

In Classical Electrodynamics by J. D. Jackson, the electric susceptibility is plotted against 1/T for polar and non-polar substances, respectively. The plot is a horizontal line for non-polar substance (see Fig.1)[14].

For non-polar substances $\quad\quad\quad \chi_e = N\alpha = N d_{int}/\varepsilon_o E = A \quad\quad\quad (15)$

where N is the number density of atoms or molecules, α is the atomic or molecular polarisability, $d_{int}$ is the induced EDM of an atom or molecule, $\varepsilon_o$ is the permittivity of free space, E is the external electric field, and the constant A is independent of the temperature T. For polar substances the plot is an oblique line.

For polar substances $\quad\quad\quad \chi_e = A + N d_0^2/3kT\varepsilon_o = A + B/T \quad\quad\quad (16)$

where $d_0$ is the permanent EDM of a molecule, k is the Boltzmann constant, and the slope $B = Nd_0^2/3k\varepsilon_o$ is constant when N keeps a fixed density[14]. **If K atom is polar and has a permanent EDM, a temperature dependence of the form $\chi_e = A + B/T$ should be expected when measuring the capacitance.**

## 4. Experimental method and result

The first experiment: involves measuring the capacitance of K vapor under the condition of the saturated vapor pressure. The experimental apparatus is a closed glass container resembling a Dewar flask in shape. Its length is $L_1 = 26.0$ cm. The external and internal diameters of the container are $D_1 = 80.8$ mm and $D_2 = 56.8$ mm. The external and internal surfaces of the container are plated with silver, shown by **a**, and **b**, respectively, in Fig.1. These two silver layers constitute the cylindrical capacitor. The thickness of the glass wall is h=1.5 mm and the separation $H_1 = 9.0$ mm. This capacitor is connected in series by two capacitors. One is called C′ and contains the K vapor of thickness $H_1$; the other is called C″ and contains the glass medium of thickness 2h. The total capacitance C is C = C′C″/(C′+C″), where C″ and C can be directly measured. The magnitude of capacitance is measured by a digital capacitance meter. The precision of the meter was 0.1 pF, the accuracy was 0.5% and the surveying voltage was V=1.2 V. When the container is empty, it is pumped to vacuum pressure P ≤$10^{-8}$ Pa for 20 h to remove impurities. We measured the total capacitance C = 50.3 pF and C″=1658 pF, and the vacuum capacitance C′$_0$=51.9 pF. Next, 5g of K material with 0.9995 purity supplied by Strem Chemicals Co. USA, was put in the container. We put the capacitor into a temperature-control stove, raise the temperature of the stove very slowly and keep it at $T_1$ =533$K$ for 4 h. We measured the two capacitances ($C_t$=2270pF, and C″=6610 pF), and the capacitance of K vapor (C′$_t$ =3457pF). The formula of saturated vapor pressure of K vapor is **P=$10^{7.183-4434.3/T}$** psi (533$K$≤T≤1033$K$), where 1 psi=6894.8 Pa [8]. We obtain the saturated pressure of K vapor as $P_1$=503.5Pa at $T_1$=533$K$. From the ideal gas law, the density of K vapor was $N_1$= $P_1/kT_1$ =6.84×$10^{22}$ m$^{-3}$. Because the surveying voltage V=1.2 V, the digital meter applied the external field only with E=V/$H_2$=1.4V/cm.

J. D. Jackson once stated that "For gases at NTP the number of molecules per cubic meter is N=2.7×$10^{25}$m$^{-3}$, so that their susceptibility should be of the order of $\chi_e$≤$10^{-3}$. Experimentally, typical values of the susceptibility are 0.00054 for air, 0.0072 for ammonia vapor, 0.007 for water vapor." [13] Note that $\chi_e$ = C′$_t$/C′$_0$–1=65.6>>$10^{-3}$ for K vapor, and its density $N_1$<< 2.7×$10^{25}$ m$^{-3}$. **So the result exceeded all physicist's expectation!**

The second experiment: involves measuring the capacitance of K vapor at various temperatures T under a fixed density. The apparatus was the same as the preceding experiment but the K vapor was at a fixed density $N_2$. To control the quantity of K vapor, the container is connected to another small container containing K material by a glass tube from the top. These two containers are slowly heated to 503$K$ in the stove for 3 h and the designed experimental container is sealed. The capacitance C of the capacitor was still measured by the digital meter and its vacuum capacitance was $C_{20}$=47.2 pF. Its length is $L_2$=23.0 cm and the plate separation is $H_2$=7.5 mm. The capacitance of K vapor has been measured at several different temperature, chosen such that the density $N_2$ of K vapor remained fixed. We obtain $\chi_e$ =A+B/T≈B/T, where the intercept A≈0 and the slope of the line B=190±4($K$). The experimental results are shown in Fig.2.

The third experiment: involves measuring the capacitance of K vapor at various voltages (V) under the fixed density $N_2$ and a fixed temperature $T_3$=303$K$[15]. The apparatus was the same as in the second experiment, $C_{30}$= $C_{20}$=47.2 pF. The measuring method is shown in Fig.3. C was the capacitor filled with K vapor to be measured and $C_d$ =520pF was used as a standard capacitor. Two signals $V_c(t)$=$V_{co}$cosωt and $V_s(t)$=$V_{so}$cosωt were measured by a two channel digital real-time oscilloscope (supply by Tektronix TDS 210 USA). The two signals had the same frequency and always the same phase at different voltages. From Fig.3, we have $(V_s-V_c)/V_c$=C/$C_d$ and $C=(V_{so}/V_{co} -1)C_d$. In this experiment, $V_{so}$ could be adjusted from zero to 800 V. When $V_1=V_{co}\leq 0.4$ V, $C_1=232$ pF ($\chi_e=3.915$) is approximately constant. With increasing voltage, the capacitance decreases gradually. When $V_2=V_{co}=400$ V, $C_2=53.0$ pF ($\chi_e=0.1229$) and approaches saturation. The $\chi_e$-V curve showed that the saturation polarization of the K vapor is obvious when $E\geq V_2/H_2=5.4\times 10^4$V/m and the field intensity at midpoint of the curve is $E_{mid}\approx 10$V/cm (see Fig.4).

**5. Theory and interpretation**   The local field acting on a molecule in a gas is almost the same as the external field **E**[14]. The electric susceptibility of a gaseous polar dielectric is[16]

$$\chi_e = N\alpha + N d_o L(a)/\varepsilon_0 E \qquad (17)$$

where $a = d_o E/kT$, $d_o$ is EDM of a molecule, $\alpha$ is the molecule polarizability. $L(a) = [(e^a + e^{-a})/(e^a - e^{-a})] - 1/a$ is called the Langevin function, it is the percentage of polar molecules or atoms lined up with the field. Note that $L(a)\approx a/3$ when $a\ll 1$, and $L(a)\approx 1$ when $a\gg 1$ [16]. Next, this equation is applied to K atoms. Because the atomic polarizability of K atoms is $\alpha=43.4\times 10^{-30}$ m$^{3}$[17], the number density of K atoms $N<7\times 10^{22}$ m$^{-3}$ and the induced susceptibility $\chi_e=A=N\alpha<3.1\times 10^{-6}$ can be neglected. In addition, the induced dipole moment of K atoms is $d_{ind}=\alpha \varepsilon_0 E$[16], because $E< 6\times 10^4$V/m in the three experiments, then $d_{ind}< 2.3\times 10^{-35}$ C.m$=1.5\times 10^{-16}$ e.cm can be neglected. From Eq. (17) we obtain $\chi_e = Nd L(a)/\varepsilon_0 E$, where d is the EDM of an K atom and N is the density of K vapor. Note that $E=V/H$ and $\varepsilon_0=C_0H/S$, we obtain the polarization equation of K atoms

$$C - C_0 = \beta L(a)/a \qquad (18)$$

where $\beta= S N d^2/kTH$ is a constant. From $a=d E/kT= dV/kTH$ we obtain **the formula of atomic** permanent **EDM**

$$\mathbf{d_{atom} = (C - C_0)V / L(a)SN} \qquad (19)$$

In order to work out **L(a)** and **a** of the first experiment, note that in the third experiment when $V_1=0.4$ V, $a_1\ll 1$ and $L(a_1)\approx a_1/3$. From Eq.(18): we obtain $C_1-C_{30}=\beta/3$ and $\beta=554.4$ pF. When $V_2=400$ V, $a_2\gg 1$ and $L(a_2)\approx 1$, we obtain $C_2-C_{30} = L(a_2)\beta/a_2$. We work out $a_2 = 94.6$, $L(a_2)=L(94.6)=0.9894$. From $\mathbf{a}= d E/kT= dV/kTH$, and $a/a_2 =VT_2H_2/T_1H_1V_2$, we obtain $\mathbf{a} = 0.134$ and $L(a)=0.0447$. $L(a)=0.0447$ means that only 4.47% of K atoms are lined up with the direction of the field in the first experiment.

Notice that we deduced Eq.(19) from the formula of the parallel-plate capacitor $\varepsilon_0=C_0 H/S$, so the cylindrical capacitor must be regarded as an equivalent parallel-plate capacitor with the plate area $S= C_0 H/\varepsilon_0$. In the first experiment the equivalent plate area $S_1= C'_0 H_1/\varepsilon_0 =5.28\times 10^{-2}$ m$^2$. Substituting the values: L(a), $S_1$, V, $N_1$ and $C-C_o = C'_t -C'_0 = (3405\pm 10)$ pF, we work out

$$d_K =(C - C_o)V / L(a) S_1 N_1 =2.531\times 10^{-29}\text{C.m}= 1.582\times 10^{-8}\text{e.cm} \qquad (20)$$

The statistical error of the measured value is $\Delta d_1/d\leq 0.12$. Considering all sources of systematic error $\Delta d_2/d\leq 0.08$, and the combined error is $\Delta d/d\leq 0.15$. We obtain

$$\mathbf{d_K=[2.53\pm 0.30(\text{stat})\pm 0.20(\text{syst})]\times 10^{-29}\text{C.m} = [1.58\pm 0.19(\text{stat})\pm 0.13(\text{syst})]\times 10^{-8}\text{e. cm}} \qquad (21)$$

From 1ev=kT, we get T=11594K. In the range of the three experiment $303K\leq T\leq 533K$, kT$\ll$1ev, so the measured capacitance change $(C'_t - C'_0)$ entirely comes from the contribution of the potassium ground state.

**6. Discussion**

**A. If K atom has a large EDM, why has not been observed by other physicists during the past few decades?**   This is an interesting question. The third experiment showed that the saturation polarization of K vapor is obvious when an external electric field $E\geq 5.4\times 10^2$V/cm. When the saturation polarization occurred, nearly all K atoms (more than 98.9 %) are lined up with the electric field, and the capacitance of K vapor $C \approx C_0$($C_0$ is the vacuum capacitance)! So only under the very weak field, $E\leq E_{mid} =10$V/cm ($E_{mid}$ is the field intensity at midpoint), the large EDM of K atom can be observed. Our experimental effect is very strong because the digital capacitance meter only with $E=V/H\leq 1.6$V/cm.

**Regrettably, nearly all scientists in this field disregard the very important problem.** The result described by S. A. Murthy et al. is a typical example [18]. Their result shows that the EDM of a Cs atom is vanishing : $d_{Cs}= (-1.8\pm 6.7(\text{stat})\pm 1.8 (\text{syst}))\times 10^{-24}$e. cm. We notice that Cs atoms are placed in an external electric field, the field intensity $E=V/H=4\times 10^3$V/cm$\gg$10V/cm. The calculation showed that a=694 and $L(a)\approx 0.9986$, nearly all the Cs atoms (more than 99.8 %) turns toward the direction of the field, and $\chi_e\approx 0$ or the capacitance of the Cs vapor is the same as vacuum! So the large EDM of Cs atom has not been observed in their experiment.

**B. If K atom has a large EDM, why the linear Stark effect has not been observed?** This is a challenging question. Let us first treat the linear Stark shifts of the hydrogen (n=2). Notice that the fine structure of the hydrogen (n=2) is only 0.33 cm$^{-1}$ for the Hα lines of the Balmer series, where $\lambda$ = 656.3 nm, and the splitting is only $\Delta\lambda$ = 0.33×(656.3×10$^{-7}$)$^2$ =0.014 nm, therefore the fine structure is difficult to observe [19]. The linear Stark shift of the energy levels is proportional to the field strength: $\Delta W = d_H E = 1.59 \times 10^{-8}$ E e.cm. When E=10$^5$V/cm, $\Delta W$=1.59×10$^{-3}$ eV, this corresponds to a wavenumber of 12.8 cm$^{-1}$. So the linear Stark shifts is $\Delta\lambda = \Delta W \lambda^2/hc$ = 12.8×(656.3×10$^{-7}$)$^2$ =0.55 nm. It is so large, in fact, that the Stark shift of the lines of the hydrogen is easily observed [19]. However, the most field strength for K atoms is Emax=5.4×10$^2$V/cm, if K atom has the EDM $d_K$=1.58×10$^{-8}$ e.cm, and the most splitting of the energy levels of K atoms $\Delta W$max= $d_K$ Emax= 8.53×10$^{-6}$ eV. This corresponds to a wavenumber of 6.87×10$^{-2}$ cm$^{-1}$. On the other hand, the observed values for a line pair of the first primary series of K atom (Z=19, n=4) are $\lambda_1$=769.90 nm and $\lambda_2$=766.49 nm[8]. So the most linear Stark shift of K atoms is only $\Delta\lambda = \Delta W (\lambda_1 + \lambda_2)^2 / 4hc$ = 0.0041nm. **It is so small, in fact, that a direct observation of the linear Stark shifts of K atom is not possible!**

Accurate measurements of the EDM of cesium, rubidium and sodium atoms in the ground state have been carried out, and similar results have been obtained [20-22].

**Acknowledgements**

This research was supported by the NSF of Guangdong Province (Grant No. 021377). The author thanks to Prof. Song-Hao Liu, Prof. Dang Mo, Prof. Xiang-You Huang, Prof. Zhen-Hua Guo, Dr. Yo-Sheng Zhang, Director Xun Chen and our colleagues Rui-Hua Zhou, Zhao Tang, Xue-Ming Yi, Engineer Yi-Quan Zhan, and Engineer Jia You for their help with this work.

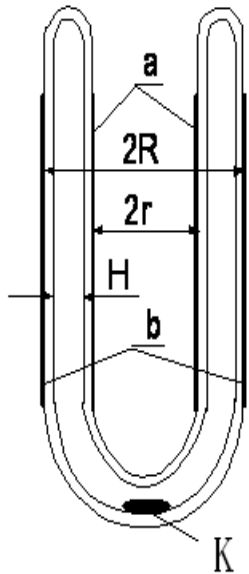

**Fig.1** This is the longitudinal section of the apparatus. It is a cylindrical glass capacitor filled with K vapor, where the density $N_1 = 6.84 \times 10^{22} m^{-3}$ when the saturated pressure of K vapor is kept at $T_1 = 533K$.

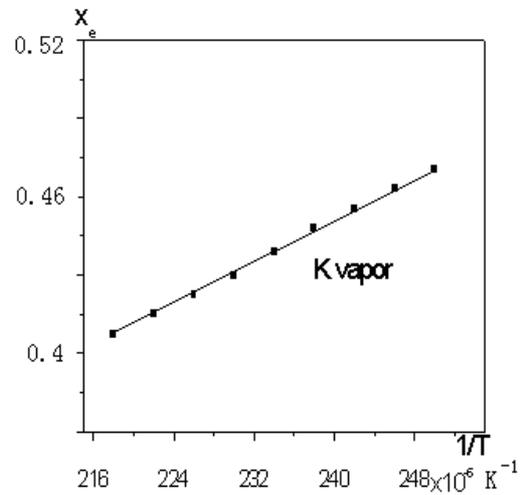

**Fig.2** The curve showed that $\chi_e$ of K vapor varies inversely proportional to the absolute temperature T: $\chi_e = A + B/T$, where the slope B=190±4.0 (*K*) and the intercept A≈0.

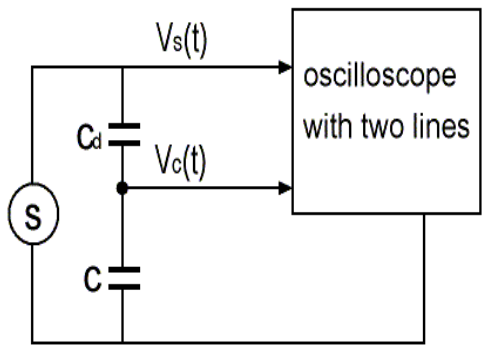

**Fig.3** The diagram shows the measuring method. C is the capacitor filled with K vapor to be measured and $C_d$=520pF is a standard capacitor, where $V_s(t) = V_{so} \cos\omega t$ and $V_c(t) = V_{co} \cos\omega t$.

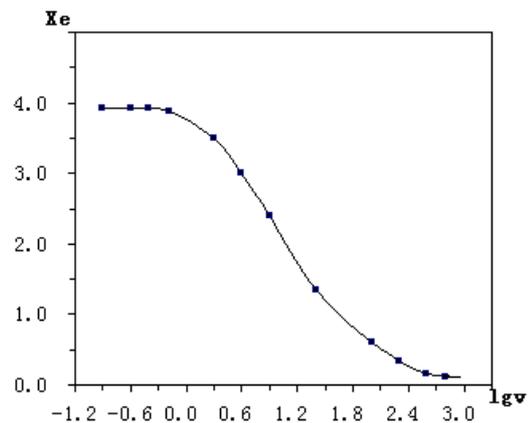

**Fig.4** The experimental curve shows that the saturation polarization effect of the K vapor is obvious when $E \geq 5.4 \times 10^2 V/cm$, and $E_{mid} \approx$ 10V/cm at midpoint of the curve.